\documentclass[aps,prl,reprint,groupedaddress]{revtex4-2}
\usepackage{graphicx}
\usepackage{amsmath}
\usepackage{amsfonts}
\usepackage{subfigure}

\begin{document}


\title{Kinetic Equality for Susceptibility and Dynamical Activity}    


\author{Takaaki Monnai}
\affiliation{Department of Science and Technology, Seikei University, Tokyo, 180-8633, Japan}


\date{\today}

\begin{abstract}
We show a general kinetic equality for susceptibilities (KSE) of general fluctuating quantities and for the dynamical activity of nonequilibrium systems described by Markovian master equations. 
As special limiting cases, KSE reproduces the so-called kinetic uncertainty relation and its multivariate extension. 
To derive KSE, we also show a kinetic generalization of fluctuation theorem, and an equality for the susceptibility and the observable-frenetic covariance. 
Hence, KSE provides a master relation of these nonequilibrium kinetic relations. 
\end{abstract}


\maketitle

 {\it Introduction.---} 
The development of nanotechnology enables us to manipulate out of equilibrium small systems such as the rectification of current and power generation in nanojunctions\cite{Hartmann1,Fujisawa2}.
Therefore, it is of fundamental and practical importance to investigate operating principles of small systems. 
This is actually one of the major subjects of the nonequilibrium statistical mechanics. 
In this context, the stochastic thermodynamics offers a powerful framework of the first and the second laws\cite{Sekimoto1,Seifert1,Seifert2} for small systems. 
In particular, general nonequilibrium relations have been derived such as fluctuation theorems (FTs)\cite{Evans1,Gallavotti1,Kurchan1,Lebowitz1,Crooks1,Jarzynski1,Gaspard1,Jarzynski2,Andrieux1,Andrieux2,Nakamura1,Rao1,Esposito1,Esposito3}, thermodynamic uncertainty relation (TUR)\cite{Seifert3,Gingurich1}, fluctuation response inequality (FRI)\cite{Dechant2,Dechant1}, and kinetic uncertainty relation (KUR)\cite{Baiesi1}. 

FT and TUR focus on the entropy production, while KUR deals with the dynamical activity, which is a frenetic quantity. 
Actually, FT provides a symmetry of the fluctuation of the entropy production, which reproduces the second law, and the linear and nonlinear response relations\cite{Lebowitz1,Andrieux1,Andrieux2}. 
TUR expresses a general upper bound of the precision of some current-like quantities in terms of the mean entropy production, which has been derived for a class of continuous time stochastic processes\cite{Gingurich1,Gingurich2,Gingurich3,Polettini1,Pietzonka1,Pietzonka2,Pietzonka3,Hasegawa5} as reviewed in \cite{Horowitz1} and subsequently extended to multidimensional cases\cite{Dechant1,Hasegawa3} and to quantum systems\cite{Goold1,Hasegawa2,Friedmann1,Monnai1}.  
Mutual relation between FT and TUR has also been intensively studied\cite{Pietzonka3,Timpanaro1,Hasegawa1,Monnai2}. For example, Ref. \cite{Monnai2} unifies the necessary and sufficient conditions of TUR and its multidimensional extensions in terms of FT.  
On the contrary, KUR gives a general upper bound of the precision in terms of the dynamical activity, which was first derived for dynamics described by Markovian master equations\cite{Baiesi1} and then extended to multidimensional cases\cite{Dechant1}. KUR can complement TUR for a certain regime of large entropy production. 
Being the statements of precision, TUR and KUR are derivable from informational theoretic approaches from the combined use of the general Cramer-Rao inequality and FRI bounding the response by Kullback-Leibler divergence (KL divergence) between probability functionals of paths of the system of interest and of a perturbed system with a specific form of perturbation\cite{Dechant2,Dechant1,Hasegawa3}.      

In this Letter, we derive a general kinetic equality for susceptibilities of arbitrary fluctuating observables and the dynamical activity, which rigorously holds even out of equilibrium for the dynamics described by Markovian master equation. 
To derive this nonequilibrium kinetic equality for susceptibilities (KSE), we also show a kinetic and informational generalization of FT and an equality for the susceptibility and covariance of the observable and a frenetic quantity. 
As special limiting cases, we show that KSE reproduces KUR and its multivariate extension. 
Note that KSE holds without the restricting conditions such as the local detailed balance and/or the initial local equilibrium. 
Hence, KSE is considered as a master equality of these nonequilibrium relations.   
\\
\\
{\it Kinetic Equality for Susceptibilities.---}
First, we describe our main result. 
Let us consider a general Markov jump process with $n$ discrete states $\{\sigma_1,\sigma_2,...,\sigma_n\}$ evolving in continuous time with the transition rates $k_{ij}$ from a state $\sigma_i$ to another state $\sigma_j$ per unit time. 
Let $\omega$ denote the trajectory from time $t=0$ to $t=\tau$, and let the probability functionals $P_0[\omega]$ and $P_\alpha[\omega]$ of the unperturbed and perturbed trajectories, respectively. Here, the subscript $\alpha$ stands for the perturbation. We specify the form of the perturbation later.  
We consider a set of observables ${\bf f}[\omega]$ given as general real-valued functionals $f_j[\omega]$ ($1\leq j\leq k$) and the logarithm of the ratio $\Sigma[\omega]=\log\left(\frac{P_0[\omega]}{P_\alpha[\omega]}\right)$. 
Note that the mean of $\Sigma[\omega]$ with respect to $P_0[\omega]$ is positive and is equal to the KL divergence $D[P_0||P_\alpha]=\sum_\omega P_0[\omega]\log\frac{P_0[\omega]}{P_\alpha[\omega]}$.  
Following Refs. \cite{Baiesi1,Dechant1,Dechant2,Hasegawa3}, we choose the specific form of perturbation as a global modification of the transition rate $k_{ij}\rightarrow k_{ij}^\alpha=(1+\alpha)k_{ij}$. This transformation accelerates the time evolution and amounts to the scaling of time\cite{Baiesi1,Dechant1,Dechant2,Hasegawa3}. We may use other choice of  analytic function $g(\alpha)$ instead of $1+\alpha$, however, the consequent result is the same up to a multiplicative constant in the limit of small perturbation  $\alpha\rightarrow 0$. 

Then, the following equality for susceptibilities and dynamical activity holds
\begin{align}
&({\bf \chi_f}(\tau),-{\cal K})\Xi^{-1}({\bf \chi_f}(\tau),-{\cal K})^{\rm T}
={\cal K}, \label{KSE1}
\end{align}
where ${\bf \chi_f}(\tau)=(\chi_1(\tau),...,\chi_k(\tau))$ denotes the set of susceptibilities $\chi_i(\tau)$ of the mean of $f_i$ ($1\leq i\leq k$) at $t=\tau$ (\ref{susceptibility1}), $\Xi$ denotes the covariance matrix of the observables ${\bf f}$ and $\frac{1}{\alpha}\Sigma$ (\ref{covariance1}), and ${\cal K}=\sum_{i\neq j}\langle n_{ij}\rangle_\tau$  stands for the dynamical activity with $n_{ij}$ being the number of transitions from the state $\sigma_i$ to $\sigma_j$ during $\tau$\cite{Baiesi1}. 
This kinetic equality for susceptibility (KSE) rigorously holds even in out of equilibrium, which is our first main result. 
The susceptibility $\chi_i(\tau)$ is defined as 
\begin{align}
&\chi_i(\tau)=\tau\frac{\partial}{\partial\tau}\langle f_i\rangle_\tau, \label{susceptibility1}
\end{align}
which is known to be equal to the susceptibility for the perturbation  $\lim_{\alpha\rightarrow 0}\frac{\partial\langle f_i\rangle_\alpha}{\partial\alpha}$ as in the case of KUR\cite{Baiesi1}. 
Here, $\langle\cdot\rangle_\tau$ denote the average at $t=\tau$ with respect to the unperturbed dynamics, while $\langle\cdot\rangle_\alpha$ denote the average at $t=\tau$ for the perturbed dynamics as in Ref. \cite{Baiesi1}. 
The matrix exponents of the covariance are defined as 
\begin{align}
&\Xi_{i,j}=\langle (f_i-\langle f_i\rangle_\tau)(f_j-\langle f_j\rangle_\tau)\rangle_\tau \; (i,j\leq k) \nonumber \\
&\Xi_{i,k+1}=\lim_{\alpha\rightarrow 0}\frac{1}{\alpha}\langle(f_i-\langle f_i\rangle_\tau)(\Sigma-\langle\Sigma\rangle)_\tau\rangle_\tau \; (i\leq k) \nonumber \\
&\Xi_{k+1,k+1}=\lim_{\alpha\rightarrow 0}\frac{1}{\alpha^2}\langle(\Sigma-\langle\Sigma\rangle_\tau)^2\rangle_\tau. \label{covariance1}
\end{align}
Comparing with KSE, the general Cramer-Rao inequality deals with $\Xi_{i,j}$ for observables ${\bf f}$ only. In other words, KSE is an equality not only for general observables ${\bf f}$ but also for the informational quantity $\Sigma$ whose average is proportional to the time reversal symmetric kinetic quantity ${\cal K}$. We will explain that contribution from such kinetic terms is not small in general even for vanishingly small $\alpha$.           
We can easily show that the matrix exponents (\ref{covariance1}) are well-defined in the limit $\alpha\rightarrow 0$. 
Here, we just mention that the scaling by $\alpha$ is necessary in (\ref{covariance1}). For example, fix $\alpha$ sufficiently small but finite, then the mean $\langle\Sigma\rangle_\tau$ is equal to KL divergence  $D[P_0||P_\alpha]=\frac{\alpha^2}{2}{\cal K}(1+{\cal O}(\alpha))$. 
Also, it is physically natural to take the parameter $\alpha$ sufficiently small in KSE as in the linear response theory.      
In what follows, we first show KUR and its multivariate extension are reproduced as special limiting cases of KSE. 
Then, we outline the derivation of KSE with some Lemmas. See supplemental material for details of the proof.    
\\
\\
{\it Corollaries}
Let us confirm that KSE actually reproduces KUR and its multivariate generalization, which is our second main result. 
For this purpose, we start from the following inequality\cite{Monnai2}. For ${}^\forall{\bf a}=(a_1,a_2,...,a_{k+1})^{\rm T}\in\mathbb{R}^{k+1}$ and an invertible  $(k+1)\times(k+1)$ positive matrix $V$
\begin{align}
&\frac{\sum_{i=1}^{k+1}a_i^2}{\sum_{i=1}^{k+1}V_{ii}}\leq{\bf a}^{\rm T}V^{-1}{\bf a} \label{inequality2}
\end{align}
holds. 
First, we consider a special case of $k=1$ and choose the observables $f_1=\lambda f$, i.e., the observable $f_1$ is redefined by multiplying an arbitrary numerical factor $\lambda$, and apply to two variables $a_1=\chi_1(\tau)$, $a_2=-{\cal K}$ with $V=\Xi$ in (\ref{inequality2}). 
Then, in the limit $\lambda\rightarrow\infty$, the left hand side is dominated by $\lambda f$ and converges to the precision  $\frac{(\tau\frac{\partial\langle f\rangle_\tau}{\partial\tau})^2}{{\rm Var}[f]}$ with the variance for unperturbed dynamics ${\rm Var}[f]$ of $f$. Together with KSE (\ref{KSE1}), we can actually derive KUR
\begin{align}
&\frac{(\tau\frac{\partial\langle f\rangle_\tau}{\partial\tau})^2}{{\rm Var}[f]}\leq{\cal K}. \label{KUR1}
\end{align}
Furthermore, we obtain a multivariate generalization of KUR by replacing $f_i$ with $\lambda f_i$ ($i\leq k$), substituting the susceptibilities $a_i=\chi_i(\tau)$ ($i\leq k$) and the dynamical activity $a_{k+1}=-{\cal K}$ in (\ref{inequality2}) in the limit $\lambda\rightarrow\infty$ 
\begin{align}
&\frac{\sum_{i=1}^k(\tau\frac{\partial\langle f_i\rangle_\tau}{\partial\tau})^2}{\sum_{i=1}^k{\rm Var}[f_i]}\leq{\cal K}. \label{KUR2}
\end{align}
In this manner, Eq. (\ref{KUR2}) provides a kinetic restriction to the precisions of observables ${\bf f}$.   
Therefore, KSE is actually a master equality of KUR.   
In the opposite limit $\lambda\rightarrow 0$, the dynamical activity is related to the variance of $\frac{\Sigma}{\alpha}$ as 
\begin{align}
&\frac{{\rm det}\Xi}{{\rm det}\tilde{\Xi}}\leq{\cal K}, \label{susceptibility3}
\end{align}
where ${\rm det}A$ denotes the determinant of $A$ and $\tilde{\Xi}$ denotes the minor of $\Xi$ deleting $k+1$-th raw and column. 
For $k=0$, (\ref{susceptibility3}) is altered by $\Xi_{k+1,k+1}={\cal K}$. 
\\
\\     
{\it Kinetic Fluctuation Theorem.---}
Once conjectured, KSE (\ref{KSE1}) can be rigorously proven on the basis of Eq. (\ref{correlation1}) as shown in Supplemental Materials. 
Here, we give more intuitive derivation by preparing underlying lemmas that lead to KSE.      
Such a natural derivation of KSE consists of several operational notions such as large deviation principle for finite duration $\tau$ by preparing identical and independent copies of the original system, and kinetic version of FT from the change of the unperturbed probability functional $P_0[\omega]$ to the perturbed one $P_\alpha[\omega]$.  

For observables $f_j[\omega]$ ($1\leq j\leq k$) and the logarithm of the ratio $\Sigma[\omega]=\log\left(\frac{P_0[\omega]}{P_\alpha[\omega]}\right)$, we can define the probability distributions of these quantities for the unperturbed and perturbed systems as 
\begin{align}
&p_0({\bf f},\Sigma) \nonumber \\
&=\int d\omega P_0[\omega]\left(\prod_{i=1}^k\delta(f_i-f_i[\omega])\right)\delta(\Sigma-\Sigma[\omega]) \label{unperturbed1}
\end{align}
and 
\begin{align}
&p_\alpha({\bf f},\Sigma) \nonumber \\
&=\int d\omega P_\alpha[\omega]\left(\prod_{i=1}^k\delta(f_i-f_i[\omega])\right)\delta(\Sigma-\Sigma[\omega]), \label{perturbed1}
\end{align}
respectively. 
Here, the integral denotes the sum over all the trajectories. 
In particular, we connect the probability distribution (\ref{unperturbed1}) to the perturbed one (\ref{perturbed1}) by a  reweighting procedure to rewrite the average for $P_0[\omega]$ as that for $P_\alpha[\omega]$ with a multiplicative weighting factor $e^{\Sigma[\omega]}$
\begin{align}
&p_0({\bf f},\Sigma)=e^\Sigma p_\alpha({\bf f},\Sigma). \label{reweighting1}
\end{align}
Eq. (\ref{reweighting1}) resembles to FT, however, $\Sigma[\omega]$ is a kinetic quantity and its average is proportional to the dynamical activity ${\cal K}$ instead of the entropy production.  
Note that FT for the entropy production can be derived by applying the reweighting procedure to the probability functionals  of time reversal twins\cite{Jarzynski1,Esposito3}.        
Therefore, (\ref{reweighting1}) can be viewed as a kinetic version of FT.     
\\
\\
{\it Large Deviation Principle.---}
Since we fix the duration $\tau$ finite, a large parameter other than time is necessary to discuss the large deviation principle\cite{Ellis1} of $p_0({\bf f},\Sigma)$. 
For this purpose, we consider the following thought experiment and introduce the number of copies $N$ as the large parameter. 
We prepare $N$ identical and independent copies of the original system \cite{Monnai2}.  
To explain the notion of copies, we explore the microscopic Hamiltonian dynamics underlying the master equation. 
The copies are identical in the sense that the quantities such as volumes, number of particles, and corresponding Hamiltonian are exactly the same. On the other hand, the time evolutions are mutually different each other due to the  uncontrolable precise of the initial state. In the Hamiltonian dynamics, the initial states of the copies are randomly sampled from a common statistical ensemble, and the subsequent time evolutions obey the same Hamiltonian.  
Let us return to the stochastic processes.  
The trajectories $\omega_n$ of the $n$-th copy ($n=1,2,3,...$)  show statistical fluctuation for dynamics described by Markovian master equation.    
Then, the sum of the quantity $f_i$ for ensemble of copies $f_{tot,i}=\sum_{n=1}^Nf_i[\omega_n]$ ($1\leq i\leq k$) obeys the large deviation principle.   
By exploring the ensemble of copies, we can extract the statistics of the original system $p_0({\bf f},\Sigma)$. 
Actually, the rate function 
\begin{align}
&I({\bf f},\Sigma) \nonumber \\
&=-\lim_{N\rightarrow\infty}\frac{1}{N}\log p_{tot}({\bf f}_{tot}=N{\bf f},\Sigma_{tot}=N\Sigma) \label{entropy1}
\end{align}
is well-defined for the probability distribution $p_{tot}({\bf f}_{tot}=N{\bf f},\Sigma_{tot}=N\Sigma)$ of the ensemble of copies. 
We can similarly define the rate function for the perturbed dynamics
\begin{align}
&I_\alpha({\bf f},\Sigma) \nonumber \\
&=-\lim_{N\rightarrow\infty}\frac{1}{N}\log p_{tot,\alpha}({\bf f}_{tot}=N{\bf f},\Sigma_{tot}=N\Sigma) \label{entropy2}
\end{align} 
by considering the probability distribution $p_{tot,\alpha}({\bf f}_{tot}=N{\bf f},\Sigma_{tot}=N\Sigma)$ for the ensemble of copies of the perturbed system.  
Then, the kinetic FT (\ref{reweighting1}) can be expressed by the rate functions as 
\begin{align}
&I({\bf f},\Sigma)=-\Sigma+I_\alpha({\bf f},\Sigma). \label{entropy3}
\end{align}
On the other hand, the rate function of the perturbed dynamics $I_\alpha({\bf f},\Sigma)\geq 0$ is equal to zero at  the mean values $f_i=\langle f_i\rangle_\alpha$ ($1\leq i\leq k$) and $\Sigma=\langle\Sigma\rangle_\alpha$.   
Thus, substituting $f_i=\langle f_i\rangle_\alpha$ and $\Sigma=\langle\Sigma\rangle_\alpha$ into (\ref{entropy3}), we obtain a general equality
\begin{align}
&I(\langle {\bf f}\rangle_\alpha,\langle\Sigma\rangle_\alpha)=-\langle\Sigma\rangle_\alpha. \label{entropy4}
\end{align}
The left hand side of (\ref{entropy4}) can be replaced by the quadratic contribution in the limit of small perturbation $\alpha\rightarrow 0$. It is nontrivial that the higher order contributions are negligible as $|\langle\Sigma\rangle_\alpha-\langle\Sigma\rangle_\tau|/\langle\Sigma\rangle_\tau$ is not necessarily small even for small $\alpha$.   
Nevertheless, we can analytically show this step by using the following exact equality for the susceptibilities and covariance between the observables $f_i$ and $\Sigma$
\begin{align}
&\chi_i(\tau)=-\lim_{\alpha\rightarrow 0}\langle(f_i-\langle f_i\rangle_\tau)\frac{\Sigma-\langle\Sigma\rangle_\tau}{\alpha}\rangle_\tau.  \label{correlation1}
\end{align}     
Eq. (\ref{correlation1}) is convincing in terms of a consistency among KSE, KUR, and the positivity $\Xi\geq 0$.
Actually, combined with Eq. (\ref{correlation1}), KUR (\ref{KUR1}) is equivalent to the positivity of the covariance matrix $\Xi\geq 0$.            
The details are described in Supplemental Materials. 

Hence, (\ref{KSE1}) is derived by dividing both hand sides by $\alpha^2$ and replacing the kinetic term $\langle\Sigma\rangle_\alpha-\langle\Sigma\rangle_\tau=-2\langle\Sigma\rangle_\tau(1+{\cal O}(\alpha))$ with scaled one  $-\frac{2}{\alpha^2}\langle\Sigma\rangle_\tau(1+{\cal O}(\alpha))$. Here, we used the property of KL divergence  $\langle\Sigma\rangle_\alpha=-\langle\Sigma\rangle_\tau(1+{\cal O}(\tau))$ which is equal to $\frac{\alpha^2}{2}{\cal K}$ for small $\alpha$. The corresponding matrix exponents of the covariance matrix of ${\bf f}$ and $\Sigma$ are properly rescaled to that of ${\bf f}$ and $\frac{1}{\alpha}\Sigma$ as in (\ref{covariance1}). 
This completes the derivation of (\ref{KSE1}). 
\\ 
\\
{\it Example.---}
KSE holds rigorously. To see its implication, we consider a concrete model. 
As an illustrative example, let us consider the stationary state of a thermoelectric junction which consists of a quantum dot in between the left and right reservoirs $R_1$ and $R_2$\cite{Esposito4}. The quantum dot has a sharply defined energy level $\epsilon$, and the reservoirs are in local equilibrium at inverse temperatures $\beta_1=\frac{1}{k_BT_1}$ and $\beta_2=\frac{1}{k_BT_2}$ and at chemical potentials $\mu_1$ and $\mu_2$, respectively. Without loss of generality, we assume $R_1$ is hot and $R_2$ is cold. 
The quantum dot may or may not occupied by an electron.  
The population dynamics of the quantum dot is described by a Markovian master equation
\begin{eqnarray}
&&\frac{d}{dt}
\begin{pmatrix}
p_0(t)\\
p_1(t)
\end{pmatrix} \nonumber \\
&=&
\begin{pmatrix}
-(k_{01,l}+k_{01,r}) & k_{10,l}+k_{10,r} \\
k_{01,l}+k_{01,r} & -(k_{10,l}+k_{10,r})
\end{pmatrix}
\begin{pmatrix}
p_0(t)\\
p_1(t)
\end{pmatrix}
\label{master1}
\end{eqnarray}
within the wide-band approximation. 
Here, $p_n(t)$ and $k_{nm,\nu}$ denote the probability for the number of the electron in the dot being $n$ ($n=0$ for empty and $n=1$ for occupied) and the transition rate from the occupation status $n$ to $m$ via the interaction with the reservoir $R_\nu$. 
The transition rates are given by $k_{01,\nu}=\frac{a_\nu}{1+e^{x_\nu}}$ and $k_{10,\nu}=\frac{a_\nu}{1+e^{-x_\nu}}$, where $a_\nu$ denotes the frequency and the Fermi-Dirac distribution depends on $x_\nu=\epsilon-\mu_\nu$.   
For concreteness, we choose an arbitrary quantity $f_1[\omega]$ in (\ref{KSE1}) as the number of particles transferred from the left to the right reservoirs during unit time. The corresponding susceptibility is equal to the number current per unit time.   
Then, we numerically verified that for all $\lambda$ the following relation for a generalized precision from KSE is actually satisfied 
\begin{align}
&\frac{\lambda^2\langle\chi_J\rangle_\tau^2+{\cal K}^2}{\lambda^2{\rm Var}[J]+{\cal K}}\leq{\cal K}. \label{variational1}
\end{align}
Indeed, we can accurately calculate the left hand sides of (\ref{variational1}) from the cummulants of $J[\omega]$.    
\begin{figure}
\center{
\includegraphics[scale=0.6]{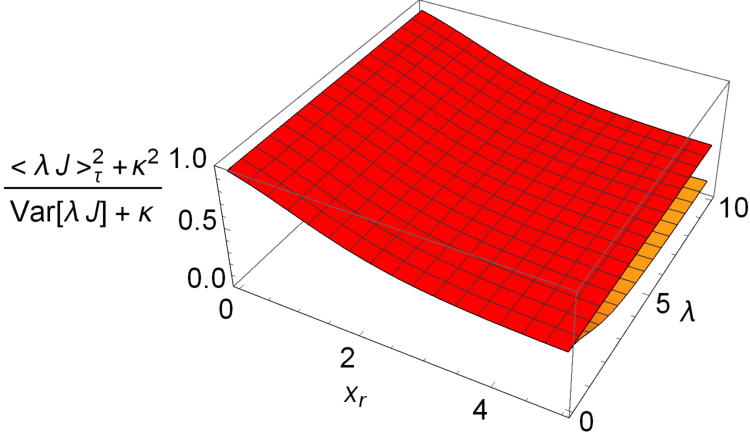}
}
\caption{Comparison of a generalized precision $\frac{\lambda^2\langle\chi_J\rangle^2+{\cal K}^2}{\lambda^2{\rm Var}[J]+{\cal K}}$ (orange) and the dynamical activity ${\cal K}$ (red) for a nanothermoelectric junction. In this case, $\langle\chi_J\rangle_\tau$ is equal to $\langle J\rangle_\tau$. We can numerically confirm that (\ref{master1}) actually holds for varying coefficient $\lambda$ and the value of $x_r=\beta_2(\epsilon-\mu_2)$. For $\lambda=0$, equality holds in (\ref{master1}). For concreteness, we fixed $a_l=a_r=1$ and $x_l=1$, and used $\lambda\in[0,10]$ and $x_r\in[0,5]$.}
\end{figure}
We can calculate the cummulants of the stationary current for Markovian master equation from full counting statistics\cite{Baiesi2}.  
In particular, the mean of the stationary particle current per unit time $J[\omega]$ obeys Landauer formula
\begin{align}
&\langle J\rangle_\tau=\frac{a_1a_2}{a_1+a_2}(f_l(x_l)-f_r(x_r)), \label{mean2}
\end{align}   
and 
the variance is given as
\begin{align}
&{\rm Var}[J]=\frac{a_1a_2}{a_1+a_2}(f_l(x_l)(1-f_r(x_r))+f_r(x_r)(1-f_l(x_l))) \nonumber \\
&-\frac{2(a_1a_2)^2}{(a_1+a_2)^3}(f_l(x_l)-f_r(x_r))^2,  \label{variance1}
\end{align}
where $f_{\nu}(x_\nu)$ stands for the Fermi-Dirac distribution of the reservoir $R_\nu$. 
On the other hand, the dynamical activity per unit time is equal to 
\begin{align}
&{\cal K}=\rho_{st,0}(k_{01,l}+k_{01,r})+\rho_{st,1}(k_{10,l}+k_{10,r}) \label{kinetic2}
\end{align}
with the stationary distribution $\rho_{st,n}$ ($n=0,1$). 
\\ 
\\
{\it Comparison with related works.---}
Let us compare KSE with preceding results.  
We derived KSE by applying Kinetic FT by reweighting the probability functional by perturbed one. 
And, the probability functionals $P_0[\omega]$ and $P_\alpha[\omega]$ differs only slightly, i.e., in the second order of perturbation.    
This point is in contrast to the case of probability functionals of paths mutually related by time reversal operation.   
If we apply the time reversal operation and FT for entropy production, TUR can be derived under geometric necessary and sufficient conditions from FT\cite{Monnai2}. 

The fact that KUR can be obtained from KSE reminds us FRI, however, KSE is qualitatively and quantitatively different because FRI provides inequality. The qualitative difference comes from the fluctuation of frenetic quantity contained in the covariance matrix of KSE. We can include the fluctuation of $\Sigma$ by using Eq. (\ref{correlation1}). Note that the difference from FRI can also be quantitatively large, since the matrix exponents $\Xi_{i,k+1}=-\chi_i$ and $\Xi_{k+1,k+1}={\cal K}$ are finite even in the limit of small perturbation $\alpha\rightarrow 0$.  
Therefore, we can not simply neglect $\langle \Sigma\rangle_\tau={\cal O}(\alpha^2)$ as small quantity. 
Mathematically, FRI can be derived if we apply a convexity inequality for the logarithm to the characteristic function $\log\langle e^{h(f-\langle f\rangle_\tau)}\rangle_\tau$ of a quantity $f$ instead of dealing with the reweighted probability distribution $p_\alpha(f,\Sigma)$. The frenetic quantity $\Sigma$ whose expectation value gives KL divergence explicitly appears in the precision of KSE while does not in FRI.   
\\ 
\\
{\it Conclusion.---}
We derived kinetic equality for the susceptibility and dynamical activity for general nonequilibirium dynamics described by Markovian master equation.  
KSE can be naturally shown by providing a direct link among the observable and the dynamical activity and a kinetic extension of FT in terms of large deviation principle. 
On the other hand, being a kinetic relation, we don't require assumptions such as the initial Gibbs ensemble and/or local detailed balance condition.    
Also, KSE reproduces nonequilibrium relations including KUR as special limiting cases.

\begin{acknowledgments}
This work was supported by the Grant-in-Aid for Scientific Research (C) (No.~18K03467 and No.~22K03456) from the Japan Society for the Promotion of Science (JSPS).
\end{acknowledgments}



\begin{thebibliography}{30}
\bibitem{Hartmann1}F. Hartmann, P. Pfeffer, S. Hofling, M. Kamp, and L. Worschech, Phys. Rev. Lett. {\bf 114}, 146805 (2015)
\bibitem{Fujisawa2}K. Chida, S. Desai, K. Nishiguchi, and A. Fujiwara, Nat. Commun. {\bf 8}, 15310 (2017)
\bibitem{Sekimoto1}K. Sekimoto, Prog. Theor. Phys. Suppl. {\bf 130}, 17(1998)
\bibitem{Seifert1}U. Seifert, Phys. Rev. Lett. {\bf 95}, 040602 (2005)
\bibitem{Seifert2}U. Seifert, Rep. Prog. Phys. {\bf 75}, 126001 (2012) 
\bibitem{Evans1}D. J. Evans, E. G. D. Cohen, and G. P. Morriss, Phys. Rev. Lett. {\bf 71}, 2401 (1993)
\bibitem{Gallavotti1}G. Gallavotti, and E. G. D. Cohen, Phys. Rev. Lett. {\bf 74}, 2694 (1995) 
\bibitem{Kurchan1}J. Kurchan, J. Phys. A {\bf 31}, 3719 (1998)
\bibitem{Lebowitz1}J. L. Lebowtiz, and H. Spohn, J. Stat. Phys. {\bf 95}, 333 (1999) 
\bibitem{Crooks1}G. E. Crooks, Phys. Rev. E {\bf 60}, 2721 (1999)
\bibitem{Jarzynski1}C. Jarzynski, J. Stat. Phys. {\bf 98}, 77-102 (2000) 
\bibitem{Gaspard1}P. Gaspard, J. Chem. Phys. {\bf 120}, 8898 (2004)
\bibitem{Jarzynski2}C. Jarzynski and D. K. Wojcik, Phys. Rev. Lett. {\bf 92}, 230602 (2004)
\bibitem{Andrieux1}D. Andrieux and P. Gaspard, J. Chem. Phys. {\bf 121}, 6167-6174 (2004) 
\bibitem{Andrieux2}D. Andrieux, P. Gaspard, T. Monnai, and S. Tasaki, New J. Phys. {\bf 11}, 043014 (2009)
\bibitem{Nakamura1}S. Nakamura, Y. Yamauchi, M. Hashisaka, K. Chida, K. Kobayashi,
T. Ono, R. Leturcq, K. Ensslin, K. Saito, Y. Utsumi, and A. C. Gossard, Phys. Rev. Lett. {\bf 104}, 080602 (2010)
\bibitem{Rao1}R. Rao, and M. Esposito, J. Chem. Phys. {\bf 149}, 245101 (2018)
\bibitem{Esposito1}M. Esposito, U. Harbola, and S. Mukamel, Rev. Mod. Phys. {\bf 81}, 1665 (2009)
\bibitem{Esposito3}M. Esposito, C. Van den Broeck, Phys. Rev. Lett. {\bf 104}, 090601 (2010)
\bibitem{Seifert3}A. C. Barato and U. Seifert, Phys. Rev. Lett. {\bf 114}, 158101 (2015) 
\bibitem{Gingurich1}T. R. Gingrich, J. M. Horowitz, N. Perunov, and J. L. England, Phys. Rev. Lett. {\bf 116}, 120601 (2016)
\bibitem{Dechant2}A. Dechant and S-I. Sasa, PNAS {\bf 117} 12 6430
\bibitem{Dechant1}A. Dechant, J. Phys. A {\bf 52}, 035001 (2019)
\bibitem{Baiesi1}I. Di. Terlizzi, M. Baiesi, J. Phys. A: Math. Theor. {\bf 52} 02LT03 (2019)
\bibitem{Gingurich2}T. R. Gingrich, G. M. Rotskoff, and J. M. Horowitz, J. Phys. A: Math. Gen. {\bf 50}, 184004 (2017)
\bibitem{Gingurich3}J. M. Horowitz and T. R. Gingrich, Phys. Rev. E {\bf 96}, 020103(R) (2017) 
\bibitem{Polettini1}M. Polettini, A. Lazarescu, and M. Esposito, Phys. Rev. E {\bf 94} 052104 (2016)
\bibitem{Pietzonka1}P. Pietzonka, A. C. Barato, and U. Seifert, Phys. Rev. E {\bf 93} 052145 (2016)
\bibitem{Pietzonka2}P. Pietzonka, A. C. Barato, and U. Seifert, J. Stat. Mech. 124004 (2016)
\bibitem{Pietzonka3}P. Pietzonka, F. Ritort, and U. Seifert, Phys. Rev. E {\bf 96}, 012101 (2017)
\bibitem{Hasegawa5}Y. Hasegawa and T. Van Vu, Phys. Rev. E {\bf 99}, 062126 (2019)
\bibitem{Horowitz1}J. M. Horowitz, and T. R. Gingrich, Nat. Phys. {\bf 16} 15 (2020)
\bibitem{Hasegawa3}T. Van. Vu and Y. Hasegawa, Phys. Rev. E {\bf 100}, 032130 (2019)
\bibitem{Goold1}G. Guarnieri, G. T. Landi, S. R. Clark, and J. Goold, 
Phys. Rev. Research {\bf 1}, 033021 (2019)
\bibitem{Hasegawa2}Y. Hasegawa, Phys. Rev. Lett., {\bf 125}, 050601 (2020)
\bibitem{Friedmann1}H. M. Friedman, B. K. Agarwalla, O. Shein-Lumbroso, O. Tal, and D. Segal, Phys. Rev. B {\bf 101}, 195423 (2020) 
\bibitem{Monnai1}T. Monnai, Phys. Rev. E {\bf 105}, 034115 (2022)
\bibitem{Timpanaro1}A. M. Timpanaro, G. Guarnieri, J. Goold, and G. T. Landi, Phys. Rev. Lett. {\bf 123}, 090604 (2019)
\bibitem{Hasegawa1}Y. Hasegawa and T. Van Vu,  Phys. Rev. Lett. {\bf 123}, 110602 (2019)
\bibitem{Monnai2}T. Monnai, Phys. Rev. E, {\bf 108} 024119 (2023)
\bibitem{Manikandan1}S. K. Manikandan, D. Gupta, and S. Krishnamurthy, Phys. Rev. Lett. {\bf 124}, 120603 (2020)
\bibitem{Hasegawa6}T. Van Vu, V. T. Vo, and Y. Hasegawa, Phys. Rev. E {\bf 101}, 042138 (2020)
\bibitem{Ellis1}R. S. Ellis, {\em Entropy, Large Deviations, and Statistical Mechanics}, Springer (1985)
\bibitem{Esposito4}M. Esposito, K. Lindenberg, and C. Van den Broeck, Europhysics Letters {\bf 105}, 60010 (2009)
\bibitem{Baiesi2}M. Baiesi, C. Maes, K. Neto\v{c}n\'{y}, J. Stat. Phys. {\bf 135}, 57-75 (2009)
\end{thebibliography}
\end{document}



\title{Supplemental Material}    


\author{Takaaki Monnai}
\affiliation{Department of Science and Technology, Seikei University, Tokyo, 180-8633, Japan}



\maketitle
\begin{widetext}
\section*{Introduction}
This Supplemental Material consists of two sections. In Sec. I, we give a rigorous proof of KSE.  
In Sec. II, we also show a mutual relation between KSE and KUR.  
\section*{I. Rigorous derivation of KSE}
Let us rigorously show KSE. 
First, we analytically calculate the difference between the quadratic approximation of the left hand side and the right hand side of Eq. (1) for single variable case $k=1$ 
\begin{align}
&(\chi_f,-{\cal K})\Xi^{-1}(\chi_f,-{\cal K})^{\rm T}-{\cal K} \nonumber \\
&=\frac{1}{D}(\Xi_{2,2}\chi_f^2+2\chi_f\Xi_{1,2}{\cal K}+{\cal K}^2\Xi_{1,1})-{\cal K} \nonumber \\
&=\frac{{\cal K}}{D}(\chi_f+\Xi_{1,2})^2. \tag{S1} \label{S1}
\end{align}
Here, $D={\rm det}\Xi$ denotes the determinant of the covariance matrix $\Xi$, and we used $\Xi_{2,2}={\cal K}$ (\ref{S3}).
Actually, we can calculate KL divergence between $P_0[\omega]$ and $P_1[\omega]=P_0[\omega]+\alpha P_1[\omega]+{\cal O}(\alpha^2)$ as 
\begin{align}
&\langle\Sigma\rangle_\tau \nonumber \\
&=\int d\omega P_0[\omega]\log\frac{P_0[\omega]}{P_\alpha[\omega]} \nonumber \\
&=-\int d\omega P_0[\omega]\log(1+\alpha\frac{P_1[\omega]}{P_0[\omega]})(1+{\cal O}(\alpha)) \nonumber \\
&=\frac{\alpha^2}{2}\int d\omega\frac{P_1[\omega]^2}{P_0[\omega]}(1+{\cal O}(\alpha)), \tag{S2} \label{S2}
\end{align}
 which is equal to $\frac{\alpha^2}{2}{\cal K}$\cite{Baiesi1}. 
Thus, $\langle\Sigma\rangle_\alpha=-\langle\Sigma\rangle_\tau(1+{\cal O}(\alpha^2))$. 
Applying $\langle\Sigma\rangle_\tau=\frac{\alpha^2}{2}{\cal K}(1+{\cal O}(\alpha))$, we can calculate $\Xi_{2,2}$ as 
\begin{align}
&\;\Xi_{2,2} \nonumber \\
&=\lim_{\alpha\rightarrow 0}\frac{1}{\alpha^2}\langle(\Sigma-\langle\Sigma\rangle_\tau)^2\rangle_\tau \nonumber \\
&=\lim_{\alpha\rightarrow 0}\frac{1}{\alpha^2}(\langle\Sigma^2\rangle_\tau-\langle\Sigma\rangle_\tau^2) \nonumber \\
&=\lim_{\alpha\rightarrow 0}\frac{1}{\alpha^2}(\int d\omega P_0[\omega](\log\frac{P_0[\omega]}{P_\alpha[\omega]})^2-\frac{1}{4}\alpha^4{\cal K}^2(1+{\cal O}(\alpha))) \nonumber \\
&=\lim_{\alpha\rightarrow 0}\frac{1}{\alpha^2}\int d\omega P_0[\omega](\log(1+\alpha\frac{P_1[\omega]}{P_0[\omega]}))^2 \nonumber \\
&=\lim_{\alpha\rightarrow 0}\frac{1}{\alpha^2}\int d\omega P_0[\omega]\alpha^2(\frac{P_1[\omega]}{P_0[\omega]})^2 \nonumber \\
&=\lim_{\alpha\rightarrow 0}\int d\omega\frac{P_1[\omega]^2}{P_0[\omega]} \nonumber \\
&={\cal K}, \tag{S3} \label{S3}
\end{align}
where we used (\ref{S2}) and $\langle\Sigma\rangle_\tau=\frac{\alpha^2}{2}{\cal K}(1+{\cal O}(\alpha))$. 

\subsection*{Fluctuation dissipation type equality for observable and dynamical activity}
We show that (\ref{S1}) is equal to zero. 
For this purpose, we calculate the covariance between $f$ and $\frac{1}{\alpha}\Sigma$
\begin{align}
&\Xi_{1,2} \nonumber \\
&=\lim_{\alpha\rightarrow 0}\frac{1}{\alpha}\langle(f-\langle f\rangle_\tau)(\Sigma-\langle\Sigma\rangle_\tau)\rangle_\tau \nonumber \\
&=\lim_{\alpha\rightarrow 0}\frac{1}{\alpha}(\langle f\Sigma\rangle_\tau-\langle f\rangle_\tau\langle\Sigma\rangle_\tau) \nonumber \\
&=\lim_{\alpha\rightarrow 0}\frac{1}{\alpha}(\int d\omega P_0[\omega]f[\omega]\log\frac{P_0[\omega]}{P_\alpha[\omega]}+{\cal O}(\alpha^2)) \nonumber \\
&=-\lim_{\alpha\rightarrow 0}\frac{1}{\alpha}\int d\omega P_0[\omega]f[\omega]\log(1+\alpha\frac{P_1[\omega]}{P_0[\omega]}) \nonumber \\
&=-\lim_{\alpha\rightarrow 0}\frac{1}{\alpha}(\int d\omega P_0[\omega]f[\omega]\frac{\alpha P_1[\omega]}{P_0[\omega]}+{\cal O}(\alpha^2)) \nonumber \\
&=\lim_{\alpha\rightarrow 0}\frac{1}{\alpha}(\langle f\rangle_\tau-\langle f\rangle_\alpha) \nonumber \\
&=-\frac{\partial\langle f\rangle_{\alpha}}{\partial\alpha}|_{\alpha=0} \nonumber \\
&=-\chi_f, \tag{S4} \label{S4}
\end{align}
which is equal to the susceptibility. 
In the fifth line, we expanded the probability functional of the perturbed dynamics as $P_\alpha[\omega]=P_0[\omega]+\alpha P_1[\omega]$ up to the first order of $\alpha$. 
Substituting (\ref{S4}) into (\ref{S1}), we obtain KSE. 
This completes the derivation for $k=1$.

\subsection*{KSE for multiple observables}
Let us show KSE for general $k$. 
The left hand side of (1) can be expanded as
\begin{align}
&({\bf \chi}_f(\tau),-{\cal K})\Xi^{-1}({\bf \chi}_f(\tau),-{\cal K})^{\rm T} \nonumber \\
&=\sum_{i=1}^k\sum_{j=1}^k\chi_i(\tau)(\Xi^{-1})_{ij}\chi_j(\tau)-2\sum_{i=1}^k\chi_i(\tau)(\Xi^{-1})_{i,k+1}{\cal K}+{\cal K}^2(\Xi^{-1})_{k+1,k+1}. \tag{S5} \label{S5}
\end{align}
From the cofactor expansion, we can evaluate each term of (\ref{S5}). 
First, we calculate a term containing $(k+1,k+1)$ matrix component  
\begin{align}
&{\cal K}^2(\Xi^{-1})_{k+1,k+1} \nonumber \\
&=\frac{{\cal K}}{D}\left|
\begin{array}{ccccc}
\Xi_{1,1} & \Xi_{1,2} & \cdots & \Xi_{1,k} & 0 \\
\Xi_{2,1} & \Xi_{2,2} & \cdots & \Xi_{2,k} & 0 \\
\vdots & \multicolumn{2}{c}{\dotfill} &\vdots &0 \\
\Xi_{k,1} & \Xi_{k,2} & \cdots & \Xi_{k,k} & 0 \\
\Xi_{k+1,1} & \Xi_{k+1,2} &\cdots &\Xi_{k,k+1} &\Xi_{k+1,k+1}
\end{array}
\right|, \tag{S6} \label{S6}
\end{align}
where $D={\rm det}\Xi$ denotes the determinant and we used $\Xi_{k+1,k+1}={\cal K}$. 
By using $\chi_i=-\Xi_{i,k+1}$ from (\ref{S4}), we can also evaluate the first and the second terms of (\ref{S5}). 
The second term is calculated as 
\begin{align}
&-2\sum_{i=1}^k\chi_i(\tau)(\Xi^{-1})_{i,k+1}{\cal K} \nonumber \\
&=\frac{2{\cal K}}{D}\left|
\begin{array}{cccccc}
\Xi_{1,1} &\cdots & \Xi_{1,i} &\cdots &\Xi_{1,k} &\Xi_{1,k+1} \\
\Xi_{2,1} &\cdots & \Xi_{2,i} &\cdots &\Xi_{2,k} &\Xi_{2,k+1} \\
\vdots &\multicolumn{4}{c}{\dotfill} &\vdots  \\
\Xi_{k,1} &\cdots &\Xi_{k,i} &\cdots &\Xi_{k,k} &\Xi_{k,k+1} \\
\Xi_{k+1,1} &\cdots&\Xi_{k,i}&\cdots &\Xi_{k+1,k}& 0
\end{array}
\right|. \tag{S7} \label{S7}
\end{align}
Similarly, the first term is evaluated as 
\begin{align} 
&\sum_{i=1}^k\sum_{j=1}^k\chi_i(\tau)(\Xi^{-1})_{i,j}\chi_j \nonumber \\
&=-\frac{{\cal K}}{D}\left|
\begin{array}{cccccc}
\Xi_{1,1} &\cdots & \Xi_{1,i} &\cdots &\Xi_{1,k} &\Xi_{1,k+1} \\
\Xi_{2,1} &\cdots & \Xi_{2,i} &\cdots &\Xi_{2,k} &\Xi_{2,k+1} \\
\vdots &\multicolumn{4}{c}{\dotfill} &\vdots  \\
\Xi_{k,1} &\cdots &\Xi_{k,i} &\cdots &\Xi_{k,k} &\Xi_{k,k+1} \\
\Xi_{k+1,1} &\cdots&\Xi_{k,i}&\cdots &\Xi_{k+1,k}& 0
\end{array}
\right|. \tag{S8} \label{S8}
\end{align} 
Substituting (\ref{S6}), (\ref{S7}), and (\ref{S8}) into (\ref{S5}),   
KSE is derived
\begin{align}
&({\bf \chi}_f(\tau),-{\cal K})\Xi^{-1}({\bf \chi}_f(\tau),-{\cal K})^{\rm T}={\cal K}. \tag{S9} \label{S9}
\end{align} 

\section*{II. Comparison with KUR}
We derive KUR from the positivity of the covariance matrix $\Xi$ and Eq. (15). 
For $k=1$, the determinant of the covariance matrix is equal to the product of eigenvalues $\lambda_1$ and $\lambda_2$
\begin{align}
&\;{\rm det}\Xi \nonumber \\
&={\rm Var}[f]{\cal K}-(\frac{\langle(f-\langle f\rangle_\tau)(\Sigma-\langle\Sigma\rangle_\tau)\rangle_\tau}{\alpha})^2  \nonumber \\
&={\rm Var}[f]{\cal K}-\chi_f^2 \nonumber \\
&=\lambda_1\lambda_2\geq 0. \tag{S10} \label{S10}
\end{align}
In the third line, we used (\ref{S4}). From the positivity of $\Xi$, the eigenvalues are positive $\lambda_j\geq 0$.  
Eq. (\ref{S10}) shows KUR.

Let us directly calculate the difference between the left hand sides of KSE and of KUR 
\begin{align}
&(\chi_f,-{\cal K})\Xi^{-1}(\chi_f,-{\cal K})^{\rm T}-\frac{\chi_f^2}{{\rm Var}[f]} \nonumber \\
&=\frac{({\rm Var}[f]{\cal K}+\Xi_{1,2}\chi_f)^2}{{\rm Var}[f]({\rm Var}[f]{\cal K}-\Xi_{1,2}^2)}\geq 0. \tag{S11} \label{S11} 
\end{align}
Eq. (\ref{S11}) is actually consistent with KSE (1) and KUR (5).


\end{widetext}